\documentclass
[twocolumn,showpacs,preprintnumbers,amsmath,amssymb,aps,pre,superscriptaddress,floatfix,reprint]
{revtex4}
\usepackage{graphicx}% Include figure files
\usepackage{dcolumn}% Align table columns on decimal point
\usepackage{bm}% bold math
\usepackage{here}
\usepackage{color}

\begin{document}

%Dynamical Anderson Transition in Kicked Models
\title{Critical Phenomena of Dynamical Delocalization in Quantum Anderson Map}
%\title{A Numerical Test of Pade Approximation for Some Functions}
\author{Hiroaki S. Yamada}
%\email{hyamada[at]uranus.dti.ne.jp}
\affiliation{Yamada Physics Research Laboratory,
Aoyama 5-7-14-205, Niigata 950-2002, Japan}
\author{Fumihiro Matsui} 
%\email{rp00863[at]ed.ritsumei.ac.jp}
\affiliation{Department of Physics, College of Science and Engineering, Ritsumeikan University
Noji-higashi 1-1-1, Kusatsu 525-8577, Japan}
\author{Kensuke S. Ikeda}
%\email{ahoo[at]ike-dyn.ritsumei.ac.jp}
%}
\affiliation{College of Science and Engineering, Ritsumeikan University
Noji-higashi 1-1-1, Kusatsu 525-8577, Japan}

\date{\today}
\begin{abstract}
Using a quantum map version of one-dimensional Anderson model, 
the localization-delocalization transition of quantum diffusion 
induced by coherent dynamical perturbation is investigated in 
comparison with quantum standard map. Existence of critical phenomena, 
which depends on the number of frequency component $M$, is demonstrated.
Diffusion exponents agree with theoretical prediction for the
transition, but the critical exponent of the localization length deviates 
from it with increase in the $M$. The critical power $\epsilon_c$ of the normalized 
perturbation at the transition point remarkably decreases as 
$\epsilon_c \sim (M-1)^{-1}$.
\end{abstract}

%05.45.Mt Quantum chaos; semiclassical methods
%71.23.Anderson localization: disordered solids
%72.20.Ee Mobility edges
%%%%%%%%%%%%%%%%%%
%%%%%%%%%%%%%%%%%%
%%%%%%%%%%%%%%%%%%
%03.00.00 Quantum mechanics, field theories, and special relativity
%03.65.-w Quantum mechanics
%03.75.-b Matter waves
%%%%%%%%%%%%%%%%%%
%05.00.00 Statistical physics, thermodynamics, and nonlinear dynamical systems
%05.60.Gg Quantum transport
%05.30.-d Quantum statistical mechanics
%%%%%%%%%%%%%%%%%%
%71.00.00 Electronic structure of bulk materials
%71.23.An Theories and models; localized states
%%%%%%%%%%%%%%%%%%
%72.00.00 Electronic transport in condensed matter
%72.10.-d Theory of electronic transport; scattering mechanisms
%72.10.Bg General formulation of transport theory
%72.10.Di Scattering by phonons, magnons, and other nonlocalized excitations
%72.20.Ee Mobility edges; hopping transport
%72.80.Ng Disordered solids
%%%%%%%%%%%%%%%%%%
%73.00.00 Electronic structure and electrical properties of surfaces, 
%interfaces, thin films, and low-dimensional structures
%73.43.Cd Theory and modeling

%\pacs{05.45.Mt,03.65.-w,05.30.-d}
%\keywords{}

\maketitle

%\section{}
%\subsection{}
%\subsubsection{}

%% \u30d9\u30af\u30c8\u30eb\u3001\u5927\u5206\u6570\u30b3\u30de\u30f3\u30c9
\newcommand{\vc}[1]{\mbox{\boldmath $#1$}}
%% \u5206\u6570
\newcommand{\fracd}[2]{\frac{\displaystyle #1}{\displaystyle #2}}
%% \u8d64\u8272
\newcommand{\red}[1]{\textcolor{red}{#1}}
\newcommand{\blue}[1]{\textcolor{blue}{#1}}
%% \u504f\u5fae\u5206
\newcommand{\del}{\partial}

\def\ni{\noindent}
\def\nn{\nonumber}
\def\bH{\begin{Huge}}
\def\eH{\end{Huge}}
\def\bL{\begin{Large}}
\def\eL{\end{Large}}
\def\bl{\begin{large}}
\def\el{\end{large}}
\def\beq{\begin{eqnarray}}
\def\eeq{\end{eqnarray}}

%%short-handed
\def\eps{\epsilon}
\def\th{\theta}
\def\del{\delta}
\def\omg{\omega}

\def\e{{\rm e}}
\def\exp{{\rm exp}}
\def\arg{{\rm arg}}
\def\Im{{\rm Im}}
\def\Re{{\rm Re}}

\def\sup{\supset}
\def\sub{\subset}
\def\a{\cap}
\def\u{\cup}
\def\bks{\backslash}

\def\ovl{\overline}
\def\unl{\underline}

\def\rar{\rightarrow}
\def\Rar{\Rightarrow}
\def\lar{\leftarrow}
\def\Lar{\Leftarrow}
\def\bar{\leftrightarrow}
\def\Bar{\Leftrightarrow}

\def\pr{\partial}

\def\Bstar{\bL $\star$ \eL}

%%%%%%%%%%%%%%%%%%%%%%%%%%%
%%%LOCALDEFLOCALDEFLOCALDEF
%%%%%%%%%%%%%%%%%%%%%%%%%%%
% LQPU
\def\etath{\eta_{th}}
\def\irrev{{\mathcal R}}
%%%%%%%%%%%%%%%%%%%%%%%%%%%
\def\e{{\rm e}}
\def\noise{n}
%\textcolor{blue}{example}
%%%LOCALDEFLOCALDEF%%%%
%%% hat %%%%
\def\hatp{\hat{p}}
\def\hatq{\hat{q}}
\def\hatU{\hat{U}}

%%%%LOCALDEFLOCALDEFLOCALDEF%%%%
\def\iset{\mathcal{I}}
\def\fset{\mathcal{F}}
\def\pr{\partial}
\def\traj{\ell}
\def\eps{\epsilon}
\def\U{\hat{U}}
%%%%%%%%%%%%%%%%%%%%%%%%%%%%%%%%

%%%%%%%%%%%%%%%%%%%%%%%%%%%%%%%
%%%%%%%%%%%%%%%%%%%%%%%%%%%%%%%
\section{Introduction and Models} 
%%%%%%%%%%%%%%%%%%%%%%%%%%%%%%%
%%%%%%%%%%%%%%%%%%%%%%%%%%%%%%%
The localization phenomena are persistent and robust 
in one-dimensional disordered systems (1DDS) \cite{ishii73,lifshiz88,stollmann01}.
It still remains even in two-dimensional disordered systems.
However, if the dimension $d$ of the disordered system is more than 2, the localization
becomes unstable, and the localization-delocalization transition (LDT) 
takes place, and finally an irreversible diffusion sets in 
when we consider quantum diffusion of an initially localized wavepacket.
The critical phenomena of LDT have been extensively studied
on the basis of the one-parameter scaling theory (OPST) of the localization
by many authors \cite{abrahams79,macKinnon81,slevin97,croy11,markos06,garcia07}.
%{\red{
The recent reviews and developments for LDT 
have been given in a commemorative book \cite{abrahams10}, 
and references therein.
%}}

On the other hand, similar localization phenomena were
discovered for the quantum kicked rotors (KR) typically exemplified by 
the quantum standard map (SM), and it can be interpreted as the localization 
phenomenon of a class of 1DDS in terms of 
Maryland transformation \cite{grempel84,casati89,ikeda93,garcia08,tian11}. 
In this context, the additional dimensionality $(d-1)$ corresponds to the
number of the dynamical degrees of freedom $M$ applied to the KR, and
the LDT in 1DDS corresponds to the ergodic transition 
when we consider the dynamically perturbed standard maps. 
Based upon this correspondence, the critical phenomenon of the LDT
was observed for Cesium atoms in optical lattice settings \cite{chabe08,rodriguez09}.

On the analogy of the standard map's case, we can expect that the
localization in the 1DDS is unstable against the dynamical
perturbations. We have proposed the delocalization scenario that the dynamical 
perturbation to 1DDS in general enhances the localization 
length and restore the diffusive motion in a strong perturbation regime \cite{yamada99}.
Considering that electrons are interacting with lattice vibrations,
the effect of dynamical perturbation by phonon modes is essential.
It models the fundamental 
{\it dynamical and deterministic } process of the quantum 
 electronic motion turning into diffusive one which allows time-irreversible 
kinetic description.
%%%%%%%%%%
   
Here, a basic question arises. {\it Whether the LDT happens     
in the 1DDS under the interaction with dynamical 
degrees of freedom, and if it happens, how the nature of LDT changes 
with increase in the mode number $M$.} 
To answer this we introduce the 1D Anderson map (AM) described 
by the unitary time-evolution operator
\begin{eqnarray}
\label{eq:tight-binding}
% \U_n=\e^{-i\Delta\cos(p/\hbar)/2\hbar} \e^{-i\Delta(f(t_n)V(q)/\hbar)} 
%\e^{-i\Delta\cos(p/\hbar)/2\hbar}  
 \U_m=\e^{-i\Delta T(p)/2\hbar} \e^{-i\Delta(f(t_m)V(q)/\hbar)} 
\e^{-i\Delta T(p)/2\hbar},
\end{eqnarray}
for wave function defined on the discrete lattice site $q(=n)$, 
where $T(p)=2\cos(p/\hbar)=\e^{-d/dq}+\e^{+d/dq}$ and 
$V(q)(=V(n))$ is random on-site potential uniformly distributed 
over the range $[-W,W]$ \cite{yamada04,yamada12}. 
The dynamical perturbation is modeled by the sinusoidal periodic
perturbation superposed onto the on-site energy as
\beq
f(t)=1+ \frac{\eps}{\sqrt{M}}\sum_{k=1}^{M} \cos(\omega_k t), 
\eeq
where $M$ and $\eps$ are the number of the frequency component and 
the strength of the perturbation, respectively. 
The time evolution by the operator $\hat{U}_m$
approximates the unitary evolution of the dynamically perturbed Anderson model,
%$i\partial_t u(n,t) = u(n-1,t)+u(n+1,t)+f(t)V(n)u(n,t), $
\beq
i \hbar \frac{\partial u(n,t)}{\partial t}  = u(n-1,t)+u(n+1,t)+f(t)V(n)u(n,t), 
\eeq
for a short time interval $\Delta$ up to the correction of $O(\Delta^3)$, and 
the unperturbed Anderson map ($\eps=0$) retains the localization properties
 of the Anderson model \cite{yamada04,yamada12}. 
In addition, the numerical verification of the presence of LDT 
is very hard for the perturbed Anderson model, therefore we examine here the 
perturbed Anderson map to explore the presence of LDT, where we take
$\Delta=1$, typically.
Note that the strength of the perturbation is divided 
by $\sqrt{M}$ so as to make the total power of long-time average independent of $M$, 
and the frequencies are taken as incommensurate numbers of $O(1)$ \cite{frequecy}.
Replacing by $T(p)=p^2/2$ and $V(q)=K\cos(q)$, Eq.(\ref{eq:tight-binding}) turns into
the SM, which exhibits the LDT in the momentum space 
\cite{casati89}. 

%In the perturbed AM both delocalized and localized dynamical
%behaviors were observed \cite{yamada04,yamada12}. 
% However, the nature of the transition from localization 
%to the delocalization has not been known, in particular,  
%for the presence of critical phenomena.
In the perturbed AM,
both localation and delocalization have been observed \cite{yamada04,yamada12}. 
However, the nature of the transition from the former
to the latter was not known, in particular, the presence 
of critical phenomena in the transition process is still unclear. 
%%%%%%%%%%
In this paper, we numerically investigate the critical 
nature of the LDT in the AM in comparison with the LDT
in the SM, which can be analyzed by the OPST \cite{chabe08,rodriguez09}.
In particular,  we are interested in the 
mode number $M$ dependence of the transition, regarding
the $M$ with the additional dimension $(d-1)$ according to 
the interpretation in the case of SM  \cite{casati89,chabe08}.  
%$M$ with the additional dimension $(d-1)$ according to 
%the case of SM  \cite{casati89,chabe08}.  
%We increase the effective dimensionality $(M+1)=d$
%beyond $3$, which will provide with a crucial test for 
%the mean-field theory of the Anderson transition.
We increase the effective dimensionality $(M+1)=d$
far beyond $3$. 
It will provide
a crucial test for the mean-field theory of the 
Anderson transition, which regards $d=3$ as the lower
bound above which the critical exponents lose the $d-$denendency.

%%%%%%%%%%%%%%%%%%%%%%%%%%%%%%%
%%%%%%%%%%%%%%%%%%%%%%%%%%%%%%%
%\section{Diffusion exponents of LDT} 
\section{Subdiffusive properties of wavepacket dynamics 
at the localization-delocalization transition} 
%%%%%%%%%%%%%%%%%%%%%%%%%%%%%%%
%%%%%%%%%%%%%%%%%%%%%%%%%%%%%%%
For $M=1$, the localization length increases exponentially 
with the coupling strengeth $\eps$,
but we could not confirm the presence of delocalized state.
However, if $M\geq 2$, we can confirm the presence of critical state, 
which evidently borders the delocalizing behavior and the localizing one. 
This will be described closely in the present section. 

%%%%%%%%%%%%%%%%%%%%%%%%%%%%%%%
\subsection{Numerical results}
\begin{figure}[htbp]
\begin{center}
\includegraphics[width=9.0cm]{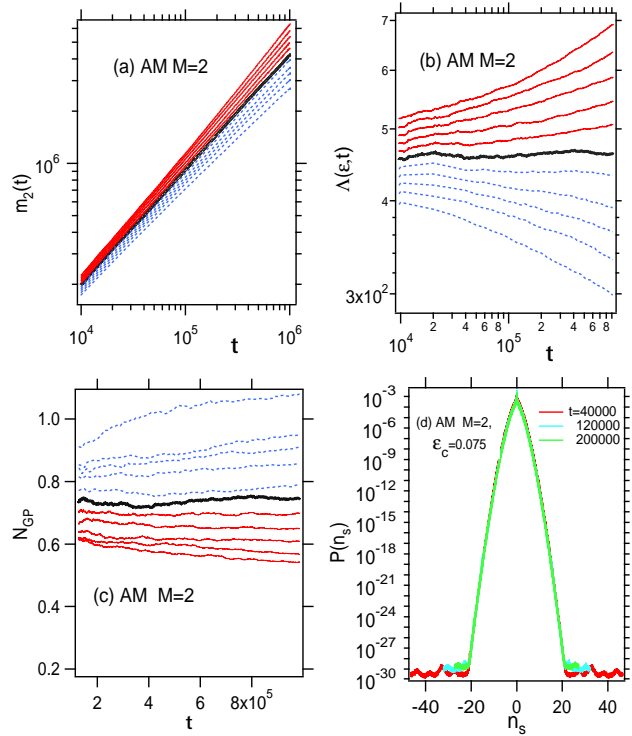}
\caption{(Color online)
The diffusive properties of the wavepacket in the perturbed AM with $M=2$.
The system and ensemble sizes are 
$N=2^{15} \sim 2^{17}$ and $10 \sim 100$, respectively,  
throughout this paper, and 
we mainly take $W=0.5$ as the disorder strength 
and $\hbar=0.125$ as the Planck constant, respectively.
(a)The double logarithmic plot of  $m_{2}(t)$ as a function of time
for different values of the perturbation strength $\eps$,
where the diffusion exponent $\alpha(=0.66)$
is determined by the least-square-fit for $m_{2}(t)$ with the critical case.
(b)The scaled MSD $\Lambda(\eps,t)$ as a function of time 
for different values of the perturbation strength $\eps$. 
Note that this is log-log plot.
(c)$N_{GP}$ as a function of time for different values of the
perturbation strength $\eps$.
Note that the black thick lines in the panels (a),(b) and (c) 
show the results at the critical case $\eps_c=0.075$.
The blue dashed curves show the results 
for $\eps<\eps_c$ in the panels (a), (b)  and (c). 
(d)Semi-log plots of the scaled probability density $P(n_s(t))$ as a function of
the $n_s=n/\sqrt{m_{2}(t)}$ 
for $t=4\times 10^{4}, 12\times 10^{4}, 20\times 10^{4}$ at the critical case.
The curves of all cases are well-overlapped.
%The vertical axis is  in the logarithmic scale.
}
\label{fig:fig1}
\end{center}
\end{figure}

%%%%%%%%%%%%%%%%%%%%%%%%%%%%%%%
Let us introduce the on-site probability $P(n,t)=|u(n,t)|^2$.
The main tool of our analysis we use the time-dependent mean square displacement (MSD) 
$m_{2}(t)= <\sum_{n=-\infty}^{\infty} (n-<n>)^2 P(n,t)>_\Omega$ of 
%%%%%%%%%%%%%%%%
%As the tool of our analysis we use the time-dependent mean square displacement (MSD) 
%$m_{2}(t)= <\sum_{n=-\infty}^{\infty} (n-<n>)^2 |u(n,t)|^2>_\Omega$ of 
the propagating wavepacket starting from localized one, $u(n,t=0)=\delta_{n,n_0}$, 
where $<\dots>_\Omega$ denotes the ensemble average over different random 
configuration of $V(n)$.
With increase in the perturbation strength, the time-evolution changes from the localized 
behavior to delocalized one passing through the critical behavior at a certain critical 
 strength $\eps=\eps_c$. (See Fig.\ref{fig:fig1}(a) for the result of AM with $M=2$. )
 At the critical state the MSD exhibits a power-law asymptotic 
(subdiffusive) dependence  $m_2(t) \sim t^\alpha$ characterized by the
diffusion exponent $\alpha$, which is close to the theoretical
value $0.66$ discussed later.
Instead of the MSD, we introduce 
the scaled MSD
\begin{eqnarray}
\Lambda(\eps,t)\equiv\frac{m_{2}(\eps,t)}{t^\alpha}
\end{eqnarray}
with respect to the critical behavior $t^\alpha$ and show its temporal evolution 
at various $\eps$ including $\eps_c$ in Fig.\ref{fig:fig1}(b).
 The critical curve indicated by the bold line  
separates the curves spreading like a fan into the delocalization regime ($\eps >\eps_c$) 
increasing up to the normal diffusion $m_2(t) \to Dt~(\Lambda \to t^{1-\alpha})$ and 
localization regime ($\eps <\eps_c$) 
decreasing down to the localization length $m_2(t) \to \xi~(\Lambda \to t^{-\alpha})$. 
Here, the diffusion constant $D=D(\eps)$ and the localization length $\xi=\xi(\eps)$ approach 
to zero and infinity, respectively, as $\eps\to\eps_c$.

\def\nst{n_s(t)}
%%%%%%%%%%%%%%%%%%%%%%%%%%%%%%%%%%%%%%%%%%
%c2-msd-w05.pxp => fig1abcd.pxp
%in c2-loc-data
%%%%%%%%%%%%%%%%%%%%%%%%%%%%%%%%%%%%%%%%%

To clarify the shape of the distribution, in addition to the MSD, we introduce
the non-Gaussian parameter (NGP) $N_{GP}$ defined by 
\beq
 N_{GP}(t) &=& \frac{1}{3} \frac{m_4(t)}{m_2(t)^2} -1, 
\eeq
where $m_4(t)=\sum  (n-<n>)^4P(n,t)$.
Figure \ref{fig:fig1}(c) depicts the time dependence of NGP 
for various $\eps$. At the critical point the NGP keeps the same nonzero-value, 
implying that the shape of the distribution function takes a similar non-Gaussian form 
throughout the time evolution \cite{gogolin76,ketzmerick97}.
%{\blue{
Figure \ref{fig:fig1}(d) shows the distribution function  
$P_s(\nst,t)=P(n,t)dn/d\nst$ at several $t$'s as a function of scaled coordinate 
$n_s(t)$  by the spread of the wavepacket for AM with $M=2$ as
\beq
 n_s(t) = \frac{n}{\sqrt{m_{2}(t)}} \propto \frac{n}{t^{\alpha/2}}.
\eeq
Evidently, the scaled representation $P_s(n_s,t)$ 
does not have explicit $t$ dependence as is expected. Thus we denote 
the scaled distribution function simply by $P_s(\nst)$.
%}}

%%%%%%%%%%%%%%%%%%%%%%%%%%%%%%%%%%%%%%%%%%
\begin{figure}[htbp]
\begin{center}
\includegraphics[width=8.5cm]{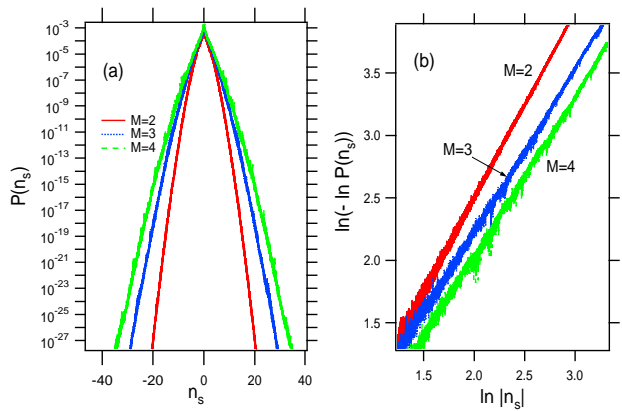}
\caption{(Color online)
(a)The scaled invariant distribution $P(n_s)$ of AM
as the function of 
$n_s=n/\sqrt{m_2(t)}$ for $M=2,3,4$
 at each critical perturbation strength $\eps_c$
from the inside to the outside.
(b)The plot of  $\ln |-\ln P(n_s)| $ as a function of 
 $|n_s|$ in the logarithmic scale.
% $\ln |n_s|$  for $M=2,3,4$
%from the top to the bottom.
The slopes correspond to the exponent $\beta$ of the
stretched Gaussian distribution.
}
\label{fig:fig2}
\end{center}
\end{figure}
%beta-prob-c2-c3a.pxp
%in c2-loc-data
%%%%%%%%%%%%%%%%%%%%%%%%%%%%%%%%%%%%%%%%%%

%{\red{ 
We further investigate the invariant function form of the wavepacket 
at each critical point of various $M$'s, as seen in Fig.\ref{fig:fig2}(a) for AM with $M=2,3,4$.
It is suggested that the tail of the scaled invariant shape of the distribution function 
 takes the stretched Gaussian distribution 
\beq
 P(n_s) \sim \exp(-|n_s(t)|^\beta),
\eeq
except for the range close to the origin of the critical state, 
where $\beta$ is the distribution exponent (streched Gaussian exponent).
%, as indicated $\beta=3/2$ for SM with $M=2$. 
The tails are shown in \ref{fig:fig2}(b) in
the plot of  $\ln |-\ln P(n_s)| $ as a function of  $\ln |n_s|$ for 
each case in Fig.\ref{fig:fig2}(a). 
The slopes of the plots correspond to the  exponents $\beta$ of the
stretched Gaussian distribution, which are decided by the least-square-fit
 except for the range close to the origin.
 The $M$-dependence of  estimated diffusion exponent $\alpha$ and
 the stretched Gaussian exponent $\beta$ are summurized in Fig.\ref{fig:fig3}.
%}}

%%%%%%%%%%%%%%%%%%%%%%%%%%%%%%%
\subsection{Comparison with theoretical prediction}
\begin{figure}[htbp]
\begin{center}
\includegraphics[width=7.0cm]{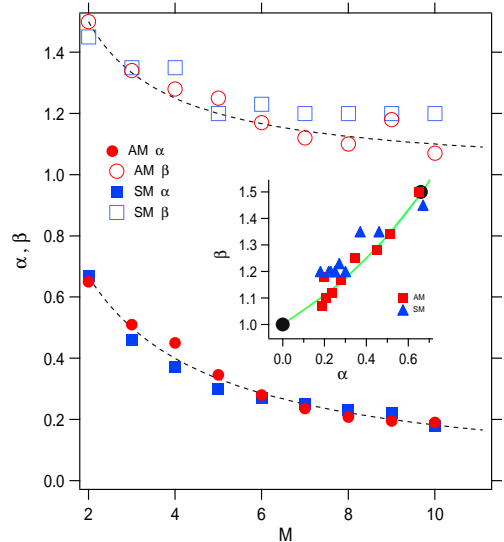}
\caption{(Color online)
The diffusion exponent $\alpha$ and distribution exponent $\beta$
as a function of $M$ for AM and SM.
The broken lines are theoretical predictions  
$\alpha_M$ and $\beta_M$, respectively.
The inset shows the plot of $\beta$ as a function of $\alpha$.
The points ($\alpha=0, \beta=1$) and ($\alpha=2/3, \beta=3/2$) 
are denoted by filled circles which  
correspond to the exponential localization and subdiffusion at $d=3$, respectively.
The green solid curve is the theoretically expected universal relation. 
The parameters  are same to the case of Fig.\ref{fig:fig1} for AM, and 
$N=2^{14} \sim 2^{16}$, $K=3.1$, $\hbar=\frac{2\pi \times311}{2^{13}}$ are for SM.
}
\label{fig:fig3}
\end{center}
\end{figure}

%%%%%%%%%%%%%%%%%%%%%%%%%%%%%%%
According to the mean-field theory of the Anderson transition in the 
$d-$dimensional disordered systems 
\cite{vollhard80}, the subdiffusion $m_2(t) \sim t^{\alpha}$ 
appears only at the critical point $\eps_c$ and the exponent is represented by the
formula $\alpha_M=\frac{2}{d}=\frac{2}{M+1}$ for $2\leq M\leq 10$. 
On the other hand, the exponent
$\beta$ has been supposed to be related with $\alpha$. For example, a phenomenological
theory based upon the assumption that the distribution function is described by a 
master equation with memory kernel tells that the relation between 
$\alpha$ and $\beta$ is an universal relation 
%$\beta=\frac{2}{(2-\alpha)}$
\beq
\beta=\frac{2}{(2-\alpha)}
\eeq
for $0 \leq \alpha \leq 1$, $1 \leq \beta \leq 2$ \cite{ketzmerick97}.
This relation predicts 
$\beta_M=(M+1)/M$ if the mean-field result is applied. The results of
our numerical experiment are compared with the theoretical prediction 
in Fig.\ref{fig:fig3}, 
and the data almost agree with the relation over a wide range $2\leq M \leq 10$. 
(The inset of Fig.\ref{fig:fig3} shows the $\alpha-\beta$ relation.)
We executed the same analysis for SM, concluding the results for AM has a great deal
in common with those of SM for $2 \leq M \leq 10$, as seen in Fig.\ref{fig:fig3}.
%%%%%%%%%%%%%%%%%%%%%%%%%%%%%%%%%%%%%%%%%%
%Fig3-new.pxp
%in c2-loc-data
%%%%%%%%%%%%%%%%%%%%%%%%%%%%%%%%%%%%%%%%%%

%{\red{
It is found that the numerical results almost agree with the theoretical predictions
except for the stretched Gaussian exponent $\beta$ in the perturbed SM with 
the larger effective dimension $(M+1) \geq 8$.
It seems that the disagreement of  $\beta$ between 
the theoretical prediction and numerical data
is due to the insufficiency of the ensemble and system size for 
the tail of the invariant distribution functions.
%It is estimated that the shortage of the sample and system size 
%cause the insufficiency of the 
%numerical results for the tail of the invariant distribution functions.
%}}

%%%%%%%%%%%%%%%%%%%%%%%%%%%%%%%
%%%%%%%%%%%%%%%%%%%%%%%%%%%%%%%
\section{Finite-time scaling analysis 
of the localization-delocalization transitions} 
\begin{figure}[htbp]   
\begin{center}
\includegraphics[width=8.5cm]{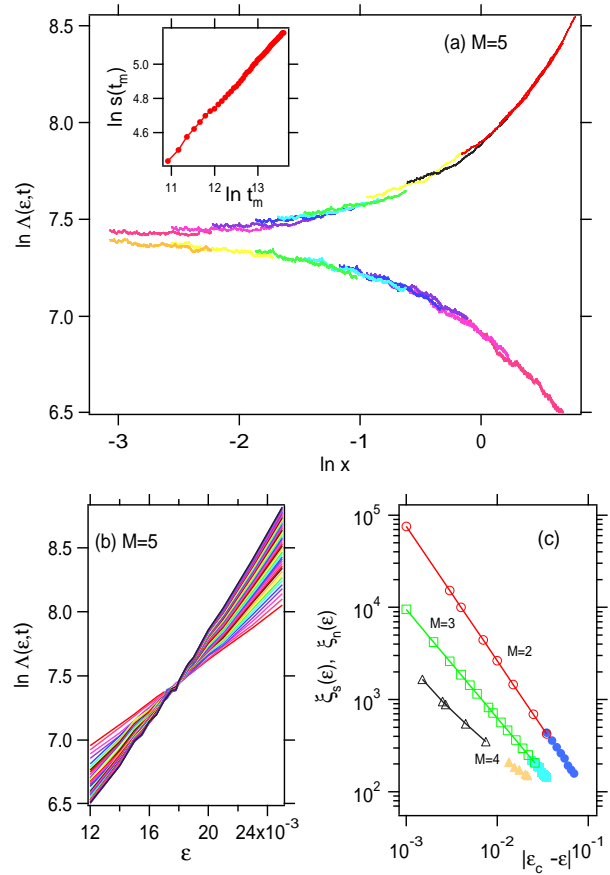}
\caption{(Color online)
The results of the critical scaling analysis for the perturbed AM.
(a)The scaled MSD $\Lambda(\eps,t)$ as a function of $x=\xi_s(\eps)t^{\alpha/2\nu}$
for some values of $\eps$ in the AM of $M=5$.
The $\eps$-dependent localization length $\xi_s(\eps)$ is determined by a scaling 
relation Eq.(\ref{eq:real-scale-0}) by least-square fit in the inset of the panel (a).
(b)The same data as panel (a) but plotted as a function of the perturbation 
strength $\eps$. In the ideal case, all lines have a common crossing point 
at $\eps_c=0.0175$.
(c)The localization length $\xi(\eps)$ as a function of 
$(\eps_c-\eps)$ for $M=2,3,4$. 
The filled symbols denote the numerical data 
directly obtained by $\sqrt{m_2(t \to \infty)}$ in the long-time limit. 
The open symbols indicate the localization length $\xi_s(\eps)$ obtained by 
OPST in the critical region. Note that the axes are in logarithmic scale.
%The filled symbols denote the numerical data that directly obtained by
%$\xi_n(\eps)=\sqrt{m_2(t \to \infty)}$  in the long-time limit. 
%The open symbols are ones  $\xi_s(\eps)$ by the numerical scaling analysis 
%with the slope $\nu$. Note that the axes are in logarithmic scale.
}
\label{fig:fig4}
\end{center}
\end{figure}

%%%%%%%%%%%%%%%%%%%%%%%%%%%%%%%
%%%%%%%%%%%%%%%%%%%%%%%%%%%%%%%
We next investigate the critical exponent $\nu$ related to 
the localization (correlation) length $\xi$, 
which is supposed to diverge $\xi \sim |\eps-\eps_c|^{-\nu}$ 
for the localized regime $\eps < \eps_c$ (for the diffusive regime $\eps > \eps_c$). 
The LDT can be in general observed both for AM and SM.
 We exhibit here the finite-time scaling analysis for AM 
taking the case of $M=5$ as the example.
First, we show in Fig.\ref{fig:fig4}(b) the observed value of $\ln \Lambda(\eps,t)$ 
at various different times $t_m$ as a function of $\eps$. A remarkable feature is
that all the curves crosses at a single point, which can be regarded as $\eps_c$.   
This fact allows us to follow the OPST which
is usually supposed for the LDT as follows:
\beq
\Lambda(\eps,t) &=& F((\eps_c-\eps)t^{\alpha/2\nu}),
\label{eq:real-scale-0}
\eeq
where $F(x)$ is a differentiable scaling function. We note that the asymptotic
function form of $F(x)$ should be $F(x) \to |x|^{-2\nu}$ in order to represent the 
localization 
$\Lambda(\eps,t) \to t^{^\alpha}x(\eps) \sim t^{^{-\alpha}}|\eps-\eps_c|^{-2\nu}$.
%Here $\xi \sim |\eps_c-\eps|^{-\nu}$ is the localization length
%(correlation length) for the $\eps < \eps_c$ ($\eps > \eps_c$).
With the above hypothesis, the scaled MSD can be expressed by 
$\ln \Lambda(\eps,t)-\ln \Lambda_c(t) \propto (\eps_c-\eps)t^{\alpha/2\nu}$
around the critical point $\eps=\eps_c$, where $\Lambda_c=F(0)$.
Applying this relation to the curves in Fig.\ref{fig:fig4}(b), we can
plot $t_m$ and the corresponding slope
$s(t_m)=\ln(\Lambda(\eps,t_m)/\Lambda_c(t_m))/(\eps_c-\eps)$ ,
 which should $\propto t^{\alpha/2\nu}$, as shown in
the inset of Fig.\ref{fig:fig4}(a), which enables to decide the
unknown exponent $\nu$ by using the known diffusion exponent $\alpha$.
With this $\nu$, we can explicitly construct the localization length function
$\xi_s(\eps)=\xi_0(\eps_c-\eps)^{-\nu}$ and further 
the scaling function
$F(\xi_s(\eps)t^{\alpha/2\nu})$ as a function of $x=\xi_s(\eps)t^{\alpha/2\nu}$.  
Figure \ref{fig:fig4}(a) 
shows the scaling functions $F(x)$ constructed by the time-dependent 
data of MSD at various different $\eps$'s close to $\eps_c$.
All the time-dependent data obtained at different $\eps$ form 
a unified function if the scaled variable $x$ is used, which 
proves numerically the validity of the OPST.

%%%%%%%%%%%%%%%%%%%%%%%%%%%%%%%%%%%%%%%%%%
%Fig4-new.pxp
%nu3-test.pxp in \u975e\u5c40\u5728\u30ce\u30fc\u30c8
%by loc-nu3.f90 in subdiffusion
%%%%%%%%%%%%%%%%%%%%%%%%%%%%%%%%%%%%%%%%%

%%%ikikik1110 少し強調的にした
%In Fig.\ref{fig:fig4}(c),  we compare  the localization length function 
%$\xi_s(\eps)$ decided indirectly by OPST which works 
%in the critical region $\eps\sim\eps_c$ with  $\xi_n(\eps)$ decided by the direct
%observation of the MSD, which can be computed numerically for various $\eps$'s 
%much less than $\eps_c$. 
In Fig.\ref{fig:fig4}(c), we compare  the localization length function 
$\xi_s(\eps)$ decided indirectly by OPST %which works 
in the critical region $\eps\sim\eps_c$ with $\xi_n(\eps)$ decided directly
by the saturated MSD data which are precisely calculatable for $\eps$'s much less 
than the critical region.
%%%%%%%%%%%%%%
The $\eps-$dependence of these two localization lengths, $\xi_s(\eps)$, $\xi_n(\eps)$,
 seem to connect continuously, which 
implies unexpected wideness of the critical region in which the OPST 
works. 
%{\red{
Accordingly,  Eq.(\ref{eq:real-scale-0}) based on the OPST immediately leads to
\beq
\label{eq:real-scale-D}  m_2(t) \propto (\eps-\eps_c)^{-2\nu} x^{2\nu}F(x),
\eeq
where $x=(\eps_c-\eps)t^{\alpha/2\nu}$ and $F(x)\to x^{-2\nu}$ for $x \to \infty$.
It describes the universal relaxation process toward the localized state, but
we do not still know to what extent the universal relaxation dynamics works
out of the critical region.
%}}

%%%%%%%%%%%%%%%%%%%%%%%%%%%%%%%
%%%%%%%%%%%%%%%%%%%%%%%%%%%%%%%
\section{Critical exponent and perturbation strength of 
the localization-delocalization transitions} %%%%%
%%%%%%%%%%%%%%%%%%%%%%%%%%%%%%%
%%%%%%%%%%%%%%%%%%%%%%%%%%%%%%%

%%%%%%%%%%%%%%%%%%%%%%%%%%%%%%%%%%%%%%%%%%
\begin{figure}[htbp]
\begin{center}
\includegraphics[width=9.0cm]{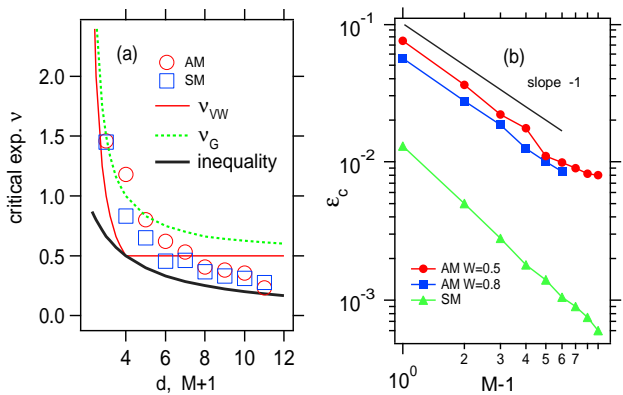}
\caption{(Color online)
(a)The dimensionality $(M+1)=d$ dependence of the critical exponent $\nu$
which characterizes the critical dynamics in the AM and SM. 
The  red solid line and green dashed line are the results of the analytical prediction by 
$\nu_{VW}$ and $\nu_{G}$, respectively.
Thick line denotes the lower bound by the Harris' critical inequality.
(b)The critical perturbation strength $\eps_c$ as a function of
$(M-1)$ for the AM and SM.
Note that the axes are in logarithmic scale.
Here, we used the SM with $K=3.1$, $\hbar=\frac{2\pi \times311}{2^{13}}$.
}
\label{fig:fig5}
\end{center}
\end{figure}
%Fig5-new.pxp
%alpha-beta-M.pxp data ha alpha-beta-M.txt + alpha-beta.txt
%by alpha-beta-1.f90
%%%%%%%%%%%%%%%%%%%%%%%%%%%%%%%%%%%%%%%%%%

In Fig.\ref{fig:fig5}(a), we compare the results of $\nu$ 
for AM and SM at various $M$ 
in comparison with theoretical predictions.
The critical exponent $\nu$ obtained from the self-consistent mean-field theory 
of the localization (VW-theory) is $\nu_{VW}=1/(d-2)$ for $2<d<4$ 
and $\nu_{VW}=1/2$ for $d \geq 4$ \cite{vollhard80}. 
The semiclassical theory of Garcia predicts $\nu_{G}=1/2+1/(d-2)$
which asymptotically approaches the value 1/2 of $\nu_{VW}$ for $d \to \infty$ \cite{garcia08b}.   
On the other hand, the inequality $\nu \geq 2/d$ is proposed at the critical point 
by Harris \cite{harris74}. Our results tell that for the larger value of  $M(\geq 7)$
the critical exponents of the AM and SM become 
significantly lower than the theoretical lowest value $1/2$ but satisfy the Harris' 
inequality \cite{harris74,chayes86,kramer93}.

%%%%%%%%%%%%%%%%%%%%%%%%%%%%%%%
%%%%%%%%%%%%%%%%%%%%%%%%%%%%%%%
%\section{Critical perturbation strength} 
%%%%%%%%%%%%%%%%%%%%%%%%%%%%%%%
%%%%%%%%%%%%%%%%%%%%%%%%%%%%%%%
Finally, we show the $M$-dependence of the critical strength $\eps_c$ 
in Fig.\ref{fig:fig5}(b).
Note that in the definition of $\eps$ we normalized the perturbation 
by $\sqrt{M}$ in order to 
make the power strength of perturbation is independent of $M$ for the fixed $\eps$. 
In spite of such a normalization, the critical perturbation strength depends
strongly upon $M$. As shown in Fig.\ref{fig:fig5}(b), our data for AM and SM indicate 
the inverse power law
\beq
\label{eq:epsc}
 \eps_c \propto (M-1)^{-\delta}, 
\eeq
up to $M=10$.  The powers are estimated as 
%%%ikikik1110有効数字はたかだかふた桁ではないか。 
%$\delta \sim 1.14$ ($W=0.5$), $\delta \sim 1.07$ ($W=0.8$) for the AM, 
$\delta \sim 1.1$, for the AM ($W=0.5$ and $0.8$), 
and $\delta \sim 1.0$ for the SM ($K=3.1$).  
%%%%%%%%%%%%%%%%%%%%%%%
Eq.(\ref{eq:epsc}) means that the total power of the perturbation, which is given 
by $M\eps_c^2$ is asymptotically proportional to $\sim 1/M$. 
Such a strong $M-$dependence is highly nontrivial
and the theoretical derivation has not been given to the best of our knowledge.
It manifests that the LDT with large $M$ is a cooperative 
phenomenon among the degrees of freedom of perturbation and 
the driven system.

It is quite interesting that the numerical data of Garcia 
and Cuevas reporting the delocalization potential threshold 
of a high-dimensional disordered tight-binding model suggests
$(d-2)^{-1}$, which seems to be closely related with our results \cite{garcia07}.

%%%%%%%%%%%%%%%%%%%%%%%%%%%%%%%
%%%%%%%%%%%%%%%%%%%%%%%%%
\section{Conclusion}  %Ikeda%Ikeda%Ikeda%Ikeda
%%%%%%%%%%%%%%%%%%%%%%%%%
%%%%%%%%%%%%%%%%%%%%%%%%%%%%%%%
We investigated critical phenomena of LDT
exhibited by polychromatically perturbed AM, which models 1DDS perturbed 
by coherent dynamical perturbations, in comparison with the SM under the same 
perturbations. We confirmed the presence of critical phenomenon for the mode number 
$M \geq 2$. The diffusion exponent $\alpha$ and distribution exponent $\beta$ agree 
well with the theoretical prediction for $M\leq 10$.
On the other hand, the critical exponent $\nu$ is significantly 
lower than the predictions of the mean-field theory for large $M$, but 
it does not violate the critical inequality. 
The critical value of normalized
perturbation strength exhibits a remarkable $M$-dependence as $\eps_c \sim (M-1)^{-1}$.
As a result, all the critical characteristics of AM agree surprisingly well with SM in spite of
their fundamental difference.
Our results open a new possibility of controlling electronic localization and conduction 
by means of externally applied stimulus implemented 
by optical and/or acoustic devices \cite{sheng95}.

%%%%%%%%%%%%%%%%%%%%%%%%%
%\red{{\it Conclusion-}} %Yamada%Yamada%Yamada%Yamada%Yamada
%%%%%%%%%%%%%%%%%%%%%%%%%
%In this work we numerically investigated the critical phenomena 
%of the localization-delocalization transition 
%exhibited by polychromatically perturbed Anderson map 
%in comparison with standard map.
%We confirmed the presence of critical phenomena for the mode number $M \geq 2$,
%and it is similar to SM in spite of
%the fundamental difference in the unperturbed model.
%The diffusion exponent $\alpha$ and distribution exponent $\beta$ agrees quite
%well with the theoretical prediction for $M\leq 10$.
%On the other hand, the critical exponent
%$\nu$ deviates lower from theoretical predictions for large $M\geq 7$
% although it do not violates the critical inequality.
%Moreover, the critical value $\eps_c$ of the normalized perturbation strength 
%shows strong $M$-dependence as $\eps_c \sim 1/(M-1)^{1.1}$.
%These results suggest that in acoustical and optical 1DDS (or quasi-1DDS)
%the localized states can be extended by increasing the mode number as a perturbation 
%and diverse LDT can be observed experimentally.  

%%%%%%%%%%%%%%%%%%%%%%%%%%%%%%%%%%%%%%%%%%%%%%%
%%%%%%%%%%%%%%%%%%%%%%%%%%%%%%%%%%%%%%%%%%%%%%%
%{\bf Acknowledgments:}
\section*{Acknowledgments}
%The authors would like to thank ///// for discussion about ////////.
%{\it Acknowledgments:}
This work is partly supported by Japanese people's tax via JPSJ KAKENHI 15H03701,
and the authors would like to acknowledge them.
They are also very grateful to Dr. T.Tsuji and  Koike memorial
house for using the facilities during this study.
%One of the authors (H.Y.) also would like to acknowledge the hospitality of 
%the Physics Division of The Nippon Dental University at Niigata
%for my stay, where part of this work was completed.

%%%%%%%%%%%%%%%%%%%%%%%%%
%%%%%%%%%%%%%%%%%%%%%%%%%


\begin{thebibliography}{00}
\bibitem{frequecy}
%{\red{
The frequencies we used are followings: 
$\omega_1=(1+\sqrt{5})/2$, $\omega_2=2\pi/\lambda$, $\omega_3=2\pi/\lambda^2$,
$\omega_4=\sqrt{3}-1$, $\omega_5=\sqrt{2}-1$, $\omega_6=\sqrt{13}/2-1$,
$\omega_7=\sqrt{11}-3$, $\omega_8=\sqrt{10}/2-1$, $\omega_9=5\sqrt{17}-20$,
$\omega_{10}=2\sqrt{19}/2-1$, 
where $\lambda$ denotes the real root of the cubic equation $x^3-x-1=0$.
We have checked for another set of the frequencies.
The whole tendency of the main result in the present paper is not depend on
the choice for the long-time calculation with large system size.
\end{thebibliography}
\end{document}